\documentclass[aps,prl,superscriptaddress,showpacs,twocolumn]{revtex4-1}
\usepackage{amssymb}
\usepackage{graphicx}
\usepackage{dcolumn}
\usepackage{bm}
\usepackage{amsmath}
\usepackage[usenames]{color}
\usepackage{textcomp}
\usepackage{subfigure}

\begin{document}

\title{Controlling higher-orbital quantum phases of ultracold atoms via
coupling to optical cavities}
\author{Hui Tan}
\thanks{These authors contributed equally.}
\affiliation{Department of Physics, National University of Defense Technology, Changsha 410073, P. R. China}
\author{Jinsen Han}
\thanks{These authors contributed equally.}
\affiliation{Department of Physics, National University of Defense Technology, Changsha 410073, P. R. China}
\author{Wei Zheng}
\email{zw8796@ustc.edu.cn}
\affiliation{Hefei National Laboratory for Physical Sciences at the Microscale and Department of Modern Physics,
University of Science and Technology of China, Hefei 230026, China}
\affiliation{CAS Center for Excellence in Quantum Information and Quantum Physics,
University of Science and Technology of China, Hefei 230026, China}
\author{Jianmin Yuan}
\affiliation{Department of Physics, Graduate School of China Academy of Engineering Physics, Beijing 100193, P. R. China}
\affiliation{Department of Physics, National University of Defense Technology, Changsha 410073, P. R. China}
\author{Yongqiang Li}
\email{li\_yq@nudt.edu.cn}
\affiliation{Department of Physics, National University of Defense Technology, Changsha 410073, P. R. China}
\affiliation{Department of Physics, Graduate School of China Academy of Engineering Physics, Beijing 100193, P. R. China}

\begin{abstract}
Orbital degree of freedom plays an important role in understanding exotic phenomena of strongly correlated materials. We study strongly correlated ultracold bosonic gases coupled to a high-finesse cavity, pumped by a blue-detuned laser in the transverse direction. Based on an extended Bose-Hubbard model with parameters adapted to recent experiments, we find that by tuning the reflection of pump laser, atoms can be selectively transferred to the odd-parity $p$-orbital, or to even-parity $d$-orbital band of a two-dimensional square lattice, accompanied with cavity-photon excitations. By interacting with cavity field, atoms self-organize to form stable higher-orbital superfluid and Mott-insulating phases with orbital-density waves, as a result of cavity induced orbital-flip processes. Our study opens the route to manipulate orbital degrees of freedom in strongly correlated quantum gases via coupling to optical cavities.
\end{abstract}

\date{\today}
\maketitle

\affiliation{Department of Physics, National University of Defense
Technology, Changsha 410073, P. R. China}

\affiliation{Department of Physics, National University of Defense
Technology, Changsha 410073, P. R. China}

\affiliation{Hefei National Laboratory for Physical Sciences at the Microscale and Department of Modern Physics,
     University of Science and Technology of China, Hefei 230026, China}
\affiliation{CAS Center for Excellence in Quantum Information and Quantum Physics,
     University of Science and Technology of China, Hefei 230026, China}

\affiliation{Department of Physics, Graduate School of China Academy of
Engineering Physics, Beijing 100193, P. R. China}

\affiliation{Department of Physics, National University of Defense
Technology, Changsha 410073, P. R. China}

%\pacs{67.85.-d, 03.75.Mn, 05.30.Jp, 05.30.Rt}

\textit{Introduction}. In condensed matters, electrons have three attributes: charge, spin and orbital. Unlike charge and spin, orbital exhibits strongly orientational properties, and plays an important role in strongly correlated materials. For instance, highly anisotropic hoppings between different orbitals lead to a so-called orbital-selective Mott transition~\cite{PhysRevLett.102.126401}, and multi-orbital involved exotic pairings may induce the debating multi-band superconductivity in heavy fermions~\cite{PhysRevLett.112.067002}. From the aspect of quantum simulations, ultracold quantum gases provide a versatile platform for simulating charge and spin degrees of freedom to investigate fundamental
condensed-matter physics problems~\cite%
{bloch2005ultracold,tokura2000orbital,lewenstein2007ultracold,RevModPhys.80.885,esslinger2010fermi}%
. However, manipulating orbital degree of freedom by using higher-Bloch bands in
optical lattices is not straightforward~\cite{wu2009unconventional,dutta2015non,li2016physics}. On one hand, fermionic atoms can
populate higher-orbital bands by Pauli principle, but that needs high
density of fermions~\cite{RevModPhys.80.885}. On another hand, bosonic atoms can be prepared in
higher-orbital bands, however, it will decay into the lowest band due to
collisions~\cite{PhysRevLett.99.200405}. Recently,
fascinating techniques have been proposed to study exotic orbital phenomena,
including shaking lattice~\cite{eckardt2017colloquium,bukov2015universal} or
bipartite-lattice setup~\cite%
{wirth2011evidence,PhysRevLett.106.015302,PhysRevLett.114.115301,PhysRevLett.126.035301,RN64}%
. Observing Fermi superfluid and strongly correlated Mott-insulating orbital
order of ultracold gases, however, is still challenging~\cite%
{will2010time,soltan2012quantum,PhysRevA.87.063638,PhysRevLett.121.265301,vargas2021orbital,p-band_Fermi}.
\begin{figure}[th!]
\includegraphics[width=0.95\linewidth]{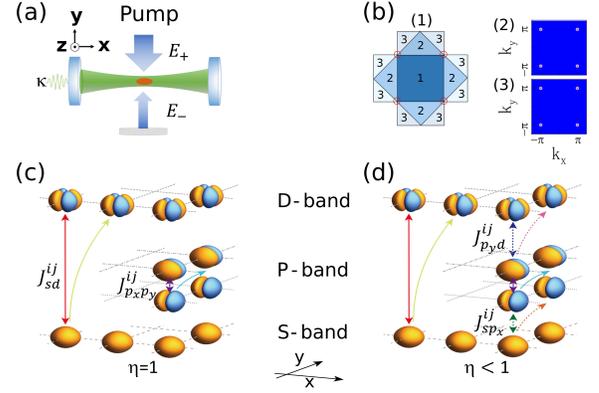}
\caption{Populating higher-orbital states with atoms in an optical cavity. (a) Atoms are prepared in an optical
cavity, pumped by a blue-detuned laser in the transverse direction with an imbalance parameter $\protect\eta= E_-/E_+$. (b) Brillouin zone of the square lattice, where atoms are scattered from the quasimomentum state $\mathbf{k}=(0,0)$
to the excite state $(\protect\pi,\protect\pi)$, with quasimomentum distributions for the $p$- and $d$-orbital bands shown in b(2) and b(3), respectively. (c)(d) Dominating scattering processes of atoms induced by cavity, leading to higher-orbital excitations. By controlling $\protect\eta$, atoms can be selectively scattered into the even-parity $d_{xy}$%
-orbital state with a single node in both $x$ and $y$ directions for $%
\protect\eta=1$ (c), or into the odd-parity $p$-orbital state with a
single node only in one direction for $\protect\eta<1$ (d). Here, $J^{ij}_{sd}$, $J^{ij}_{p_xp_y}$, $J^{ij}_{sp_x}$, and $J^{ij}_{p_yd}$ denote cavity induced orbital-flip hoppings between sites $i$ and $j$ for the $s$- and $d_{xy}$-orbitals, $p_x$- and $p_y$-orbitals, $s$- and $p_x$-orbitals, and $p_y$- and $d_{xy}$-orbitals, respectively.}
\label{schematic}
\end{figure}

Coupling ultracold atoms to a high-finesse optical cavity provides another
tool for studying quantum many-body physics~\cite%
{RevModPhys.85.553,mivehvar2021cavity}. By choosing the pump-laser frequency
smaller than the atomic internal transition (red detuning), self-organized
superradiant phase has been theoretically predicted~\cite%
{PhysRevLett.89.253003, PhysRevLett.110.113606, PhysRevLett.111.055702,
PhysRevLett.112.143002, PhysRevLett.112.143003,
PhysRevLett.112.143004,PhysRevLett.114.173903,PhysRevA.87.051604,PhysRevLett.114.123601}
and experimentally observed~\cite{2010Nature, PhysRevLett.113.070404,
2016Nature,
PhysRevLett.115.230403,PhysRevLett.120.223602,PhysRevLett.121.163601,norcia2018cavity,kroeze2019dynamical}%
, where atoms break translational symmetry by forming a density-wave
pattern, induced by cavity mediated long-range interactions. For a quantum
gas coupled to a blue-detuned cavity, self-organization of atoms should be
prohibited, since the buildup of additional repulsive potential costs energy. Surprisingly, a blue-detuned self-organized phase predicted recently~\cite%
{PhysRevLett.105.043001,PhysRevA.83.033601,
PhysRevA.85.013817,PhysRevLett.115.163601,PhysRevLett.118.073602,
PhysRevA.99.053605, Ke_ler_2020,
PhysRevA.101.061602,PhysRevResearch.3.013173}, has already been observed experimentally, and $p$-band excitation is proposed to explain this phenomenon in the weakly interacting atomic system~\cite{PhysRevLett.123.233601,PhysRevResearch.3.L012024}. A remaining open question is to identify possibilities for quantum engineering of exotic multi-orbital physics with cavity scenarios, including previously unrealized strongly correlated higher-orbital phenomena.

%A remaining open question is to identify possibilities for simulating previously unrealized strongly correlated higher-orbital phenomena of ultracold atoms via coupling to optical cavities.
%It turns out that the blue-detuned cavity scatters atoms into {\color{green}the $p$-orbital band}, , which triggers quantum engineering of exotic multi-orbital physics with cavity setups

In this letter, we investigate the collective coupling between blue-detuned cavity and atoms in a two-dimensional (2D) optical lattice. Within this 2D setup, we notice that the cavity induced scattering involves more than odd-parity $p$-orbital excitations of atoms, but also even-parity $d$-orbital population, leading to previously untouched $p$- and $d$-orbital many-body phases. As a result of cavity induced orbital-flip hopping, center-of-mass motion and orbital degree-of-freedom of atoms are coupled together, resulting in an "orbital-density wave" order in both superfluid and Mott-insulating phases. In addition, we find that populations of atoms can be selectively tuned between the $p$- and $d$-orbital bands, by controlling the reflection rate of the pump laser.

%present an experimentally related scheme to realize higher-orbital quantum phases of ultracold bosonic gases in an optical cavity, pumped by a blue-detuned laser Within this cavity setup, a considerable number of atoms can be stabilized in higher-orbital bands of the 2D square lattice due to cavity induced orbital-flipping processes,

\textit{Model and Method}. Our 2D setup is exhibited in Fig.~\ref{schematic}(a), where $^{87}$Rb atoms are loaded into a high-finesse single-mode optical cavity with a decay rate of $\kappa =40\,\omega _{r}$%
, with $\omega _{r}$ being recoil frequency. Atoms are pumped by two
counter-propagating blue-detuned lasers with wavelength $\lambda _{p}=780.1\,%
\mathrm{nm}$ in the $y$ direction perpendicular to the cavity mode, which can be
realized by applying one laser beam and reflection by a mirror. The
reflection rate controls the imbalance of the counter-propagating laser
beams $\eta \equiv E_{-}/E_{+}$~\cite{PhysRevResearch.3.L012024}, with $%
E_{+} $ and $E_{-}$ {being the electrical field amplitudes of the incident
and reflected pump lasers}, respectively. In the third direction, we assume a strong confinement freezing motional degree of freedom of atoms. In a sufficiently deep lattice, we can use the tight-binding approximation,
and keep finite relevant bands, such that the system can be described by a
generalized Bose-Hubbard model~\cite{SM}%
\begin{eqnarray}\label{Hamil}
\hat{H} &=&-\sum\limits_{\langle ij\rangle,\sigma }J_{\sigma \sigma }^{ij}\hat{b}_{i,\sigma
}^{\dag }\hat{b}_{j,\sigma }-\sum\limits_{i,\sigma }\mu_{\sigma}\hat{b}_{i,\sigma
}^{\dag }\hat{b}_{j,\sigma }-\hbar \Delta _{c}\hat{a}^{\dagger }\hat{a}   \\
&+&\sum\limits_{i,\sigma _{1}\sigma _{2}\sigma _{3}\sigma_{4}}\frac{%
U_{\sigma _{1}\sigma _{2}\sigma _{3}\sigma _{4}}}{2}\hat{b}_{i,\sigma_{1}}^{\dag }\hat{b}_{i,\sigma _{2}}^{\dag }\hat{b}_{i,\sigma _{3}}\hat{b}%
_{i,\sigma _{4}} + \hat{%
V}_{1}+\hat{V}_{2}, \notag
\end{eqnarray}%
where $\hat{V}_{1}=\frac{1+\eta}{2}(\hat{a}+\hat{a}%
^{\dagger })\sum\nolimits_{ij}\left( -1\right) ^{i}(J_{sd}^{ij}\hat{b}%
_{i,s}^{\dag }\hat{b}_{j,d}+J_{p_{x}p_{y}}^{ij}\hat{b}_{i,p_{x}}^{\dag }\hat{%
b}_{j,p_{y}}+\mathrm{H.c.})$, and $\hat{V}_{2}=-i\frac{1-\eta}{2}(\hat{a}-\hat{a}^{\dagger
})\sum\nolimits_{ij}\left( -1\right) ^{i}(J_{sp_{x}}^{ij}\hat{b}_{i,s}^{\dag
}\hat{b}_{j,p_{x}}+J_{p_{y}d}^{ij}\hat{b}_{i,p_{y}}^{\dag }\hat{b}%
_{j,d}+\mathrm{H.c.})$ are the cavity induced scattering processes, $\langle i,j\rangle$ denotes the nearest-neighbor sites, $J_{\sigma _{1}\sigma _{2}}^{ij}$ the{\ onsite (}$i=j${%
) and nearest-neighbor (}$i\neq j${) single-particle hopping amplitudes}, $\mu_\sigma \equiv J^{ii}_{\sigma\sigma}$ the chemical potential,  $\Delta _{c}$ the cavity detuning, and $%
U_{\sigma _{1}\sigma _{2}\sigma _{3}\sigma _{4}}$ the onsite interactions. $\hat{a}$ is the annihilation operator of a cavity photon, and $\hat{b}_{i,\sigma }$ the annihilation operator for the Wannier state $\sigma $ at site $i$, with $\sigma $ denoting $s$-orbital, $p_{x}$- and $p_y$-orbitals with a single node only in one direction, and $d_{xy}$-orbital with a single node in both directions, respectively.
All the Hubbard parameters are obtained from band-structure calculations of the 2D square lattice~\cite{SM}. To validate the tight-binding model, an external optical lattice with an identical wavelength of the cavity mode and a depth of $5\,E_{r}$ is added in the cavity direction, where $E_{r}$ denotes the recoil energy.
%$=\frac{h^{2}}{2m\lambda _{p}^{2}}% =\hbar \omega _{r}$ is the recoil energy, with $m$ being the mass of the $% ^{87}\mathrm{Rb}$ atom.

Generally, the dominating processes are the cavity induced scattering of atoms. Depending on parity of the scattering process, atoms can be scattered to different orbitals, controlled by atom-pump detuning~\cite{2010Nature,PhysRevLett.123.233601}. For the blue-detuned lattice system considered here, we notice that atoms can populate in both odd-parity $p$-orbital and even-parity $d$-orbital bands. As shown in Fig.~\ref{schematic}(c), $\hat{V}_{1}$ scatters atoms from the $s$- to $d_{xy}$-orbital, and from the $p_{x}$- to $p_{y}$-orbital, since $\hat{V%
}_{1}$ is associated with ${\rm cos}(k_c x){\rm cos}(k_p y)$~\cite{SM}, which is parity odd in both $x$ and $y$ directions for the blue-detuned case and changes the parity of orbitals in both $x$ and $y$ directions. Here, we choose the wave vectors of the pumping and cavity field to be identical with $k_{p}=k_{c}$. $%
\hat{V}_{2}$ scatters atoms from the $s$- to $p_{x}$-orbital, and from the $p_{y}$- to $d_{xy}$-orbital, since $\hat{V}_{1}$ is inherited from ${\rm cos}(k_c x){\rm sin}(k_p y)$~\cite{SM}, which is parity odd in the $x$ direction but even in the $y$ direction and changes the parity of orbitals only in the $x$ direction [Fig.~\ref{schematic}(d)].
In addition, due to the factor $\left( -1\right) ^{i}$, we note that both $\hat{V}_{1}$ and $\hat{V}_{2}$ transfer a quasi-momentum by $\left( \pi ,\pi \right) $ as flipping the orbitals [Fig.~\ref{schematic}(b)].

In the superradiant phase, the cavity mode is macroscopically populated, such that cavity field can be approximated by {its} mean value $\alpha (t)=\langle \hat{a}(t)\rangle$~\cite{PhysRevLett.107.140402}. In the steady state, $\partial _{t}\alpha (t)=0$, cavity field is determined self-consistently with $\alpha =\sum_{i}(-1)^{i}\langle (1+\eta) (J^{ii}_{sd}%
\hat{b}_{i,s}^{\dagger }\hat{b}_{i,d}+J^{ii}_{p_{x}p_{y}}\hat{b}%
_{i,p_{x}}^{\dagger }\hat{b}_{i,p_{y}}+\mathrm{H.c.})+i(1-\eta)(J^{ii}_{sp_{x}}\hat{b}%
_{i,s}^{\dagger }\hat{b}_{i,p_{x}}+J^{ii}_{p_{y}d}\hat{b}_{i,p_{y}}^{\dagger }%
\hat{b}_{i,d}+\mathrm{H.c.})\rangle /2( \Delta _{c}-\sum_{i,\sigma
}J_{\sigma }\langle \hat{b}_{i,\sigma }^{\dagger }\hat{b}_{i,\sigma }\rangle
+i\kappa) $, where $J_{\sigma }$ denotes the onsite matrix elements associated with the cavity mode~\cite{SM}.

\begin{figure}[tbp]
\includegraphics[width=1.0\linewidth]{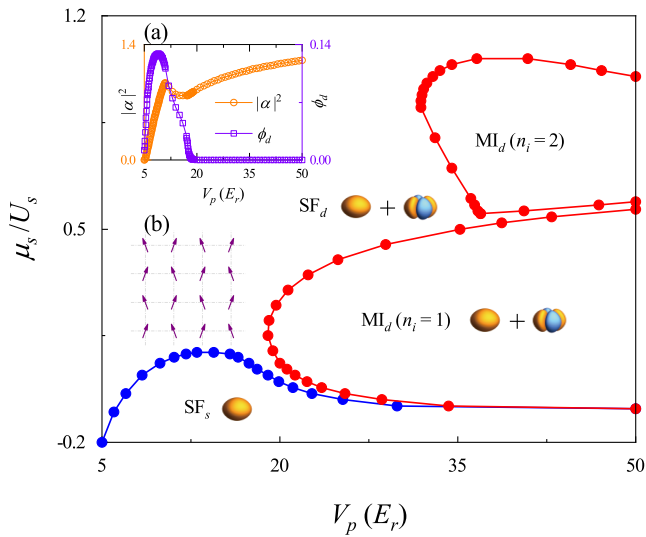}
\caption{Phase diagram of bosonic gases trapped in an optical cavity, pumped by a blue-detuned laser with an imbalance parameter $\protect\eta =1$, obtained from bosonic dynamical mean-field theory. There are three many-body phases, including the $s$-orbital superfluid phase (SF$_{s}$) without superradiance, and $d$-orbital superfluid (SF$_{d}$) and Mott-insulating (MI$_{d}$) phases with superradiance of cavity field. Inset: (a) photon number $|\protect\alpha |^2$ and $d$-orbital order parameter $\protect\phi _{d}$ are shown as a function of the pump laser depth for a fixed chemical potential $\protect\mu _{s}/U_{s}=0.17$ with $U_{s}$ being onsite interactions between atoms in the $s$-orbital band, indicating a $d$-orbital superfluid-Mott-insulating phase transition, and
(b) real-space distribution of orbital order $\langle \mathbf{\hat{S}}%
_{x,z}^{sd}\rangle $ for the $d$-orbital phases with $\langle \mathbf{\hat{S}%
}_{y}^{sd}\rangle =0$. Other parameters are $N_{\mathrm{lat}}\times
U_{0}=600\,E_{r}$, $\Delta _{c}=80\,\protect\omega _{r}$, and $\protect%
\kappa =40\,\protect\omega _{r}$, with $N_{\mathrm{lat}}$ being the total number of lattice sites and $U_0$ the light shift per photon.}
\label{new_mu_vp}
\end{figure}
To obtain the steady state of the many-body system, we numerically solve Eq.~(\ref{Hamil}) in the coherent state approximation for the cavity mode by using real-space bosonic dynamical mean-field theory (DMFT), which provides a non-perturbative description of many-body systems both in three and two dimensions~%
\cite{Vollhardt, Hubener, Werner,Li2011}, whose reliability has been compared against the quantum Monte-Carlo simulations~\cite{QMC_boson}. Recently, a four-component bosonic DMFT has been developed to study multi-species bosons in the $p$-orbital band~\cite{PhysRevLett.121.093401}.
Here, we implement this method to tackle the multi-band system, and the technical details are described in the Supplementary Material~%
\cite{SM}.

\textit{$D$-orbital population for a perfect reflection}. We first discuss
the physics for a perfect reflection of the pump laser, $\eta =1$. In this situation, ${\hat V}_2=0$, and the dominated scattering process is ${\hat V}_1$. To
characterize the many-body phases, mean cavity photon number $|\alpha |^2$,
superfluid order parameter $\phi _{\sigma }=\sum_{i}|\langle \hat{b}%
_{i,\sigma }\rangle |/N_{\mathrm{lat}}$, and orbital magnetism $\mathbf{%
\hat{S}}_{i}^{\sigma _{2}\sigma _{1}}= \hat{b}%
_{i,\sigma _{2}}^{\dagger }\mathbf{F}_{\sigma _{2}\sigma _{1}}\hat{b}%
_{i,\sigma _{1}} $ are utilized, where $N_{\mathrm{lat}}$
denotes the total number of lattice sites~\cite{lattice_size}, and $F_{\sigma _{2}\sigma _{1}}$
is the Pauli {matrices}. Since the pump laser globally couples to all atoms in the cavity, thus the atom number in turn shifts the phase boundary. This motivates us to fix the rescaled atom-cavity coupling $N_{\rm lat}\times U_0$ in our simulations, where $U_0$ is the light shift of a single atom.

\begin{figure}[tbp]
\includegraphics[width=1\linewidth]{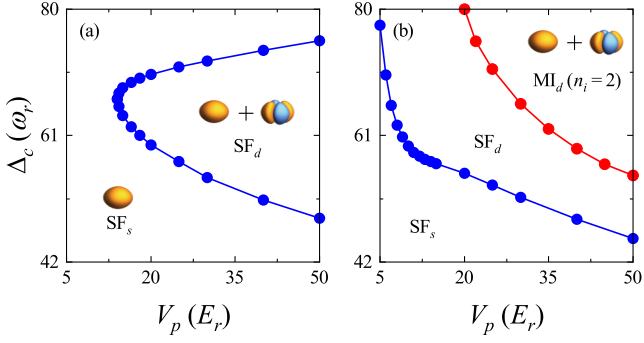}
\caption{Phase diagram of bosonic gases pumped by a blue-detuned laser with an imbalance parameter $\protect\eta=1$ as a function of cavity detuning $\Delta_c$ and pump lattice depth $V_p$, obtained from bosonic dynamical mean-field theory. (a) For the weak atom-cavity coupling $N_{\mathrm{lat}}\times U_0=240\,E_r$, there exists $s$-orbital superfluid (SF$_s$) and $d$%
-orbital superfluid (SF$_d$) phases, and (b) for the strong coupling $N_{%
\mathrm{lat}}\times U_0=400\,E_r$, there exists  $s$%
-orbital superfluid (SF$_s$), $d$-orbital superfluid (SF$_d$) and Mott-insulating (MI$_d$) phases. Here, local total filling $%
n_i=2 $, and decay rate $\protect\kappa=40\, \protect\omega_r$. }
\label{new_detuning_vp}
\end{figure}

In contrast to the red-detuned case~\cite%
{PhysRevA.87.051604}, we observe that a few percent of atoms are transferred from the $s$- to $d_{xy}$-orbital band with considerable cavity-photon excitations [Fig.~\ref{new_mu_vp}(a)], by a blue-detuned pump laser, stabilizing $d$-orbital superfluid and Mott-insulating phases. As shown in Fig.~\ref{new_mu_vp} and~\ref{new_detuning_vp}, three quantum phases appear,
including $s$-orbital superfluid phase (SF$_{s}$) with $|\alpha |^2=0$, $d$-orbital superfluid phase (SF$_{d}$) with $|\alpha |^2\neq 0$ and $\phi
_{d}\neq 0$, and $d$-orbital Mott-insulating phase (MI$_{d}$) with $|\alpha|^2\neq 0$ and $\phi _{s}=\phi _{d}=0$. As a result of the symmetry of $p$-orbital states, we only observe neglected $p$-orbital population, induced by onsite interactions~\cite{will2010time,soltan2012quantum,hazzard2010site,dutta2011bose,luhmann2012multi,PhysRevA.86.023617}. We remark here that even more higher-orbital states can be included in our numerical simulations, which is not expected to affect our results quantitatively, as a result of the large band gap in the deep lattice.

Filling-dependent phase diagram is shown in Fig.~\ref{new_mu_vp}, as a function of chemical potential and pumping strength. For a smaller chemical potential $\mu_s$, the coupling between atoms and cavity mode is so weak that the photon number in the cavity mode $|\alpha|^2=0$, with only the $s$-orbital band being populated. By raising the chemical potential,
the $d$-orbital superfluid phase appears, since the scattering is a collective effect due to all the atoms in the cavity and depends on the total particle number. For a stronger pumping strength, more photons are scattered into
the cavity mode, and the resulting standing wave in the
cavity direction suppresses tunneling of atoms with the absence of superfluidity $\phi_s=\phi_d=0$. As shown in Fig.~\ref{new_mu_vp}(a), we clearly observe the $d$-orbital superfluid-Mott-insulating phase transition upon increasing the depth of the pump laser.

Distinct from the emergent superradiant phases with charge-density waves for a red-detuned cavity, the excited atoms appear with self-organized orbital-density waves for the blue-detuned case, due to cavity induced orbital-flip hoppings, as shown in Fig.~\ref{new_mu_vp}(b), where the local total filling is homogeneous. To explain the self-organized orbital-density wave, we utilize a single-site model in the deep $d$-orbital Mott-insulating regime~\cite{SM}, and find the ground state for site $i$ to be like $|\psi _{i}\rangle
\sim (-1)^{i+1}\frac{\mu_d-\mu_s+A_1}{\sqrt{(\mu_d-\mu_s+A_1)^2+4{J_1}^2}}\vert s \rangle+\frac{2J_1}{\sqrt{(\mu_d-\mu_s+A_1)^2+4{J_1}^2}}\vert d \rangle $ {%
for filling $n_{i}=1$}, and $|\psi _{i}\rangle \sim -\frac{\mu_s-\ \mu_d+2U_{sd}-D_2}{B_2}\vert s, s \rangle+(-1)^{i+1}\frac{2J_1}{B_2}\vert s, d \rangle+\frac{\mu_s-\mu_d+D_2}{B_2}\vert d, d \rangle
$ for filling $n_{i}=2$, with the states $|s\rangle={\hat b}^\dagger_{i,s}|0\rangle$ and $|d\rangle={\hat b}^\dagger_{i,d_{xy}}|0\rangle$ for site $i$, $J_1=Re[\alpha](1+\eta)J^{ii}_{sd}$, and $A_1$, $B_2$ and $D_2$ determined by Hubbard parameters~\cite{SM}, indicating an orbital-density wave with $\langle
\hat{S}_{x}^{sd}\rangle _{i}=-\langle \hat{S}_{x}^{sd}\rangle _{i+1}$ for filling $n_{i}=1$ and $2$. These analytical results are consistent with our numerical simulations.

To be relevant to the experiments, we also map out phase diagrams as a function of cavity detuning $\Delta _{c}$ and pump lattice depth $V_{p}$ for a
fixed filling $n_{i}=\sum_{\sigma }\langle \hat{b}_{i,\sigma }^{\dagger }%
\hat{b}_{i,\sigma }\rangle =2$, {\
with $N_{\mathrm{lat}}\times U_{0}=240\,E_{r}$ (Fig.~\ref{new_detuning_vp}a), and $N_{\mathrm{lat}%
}\times U_{0}=400\,E_{r}$ (Fig.~\ref{new_detuning_vp}b)}, respectively. Three distinct phases exist for the
parameters studied here, including the SF$_{s}$, SF$_{d}$ and MI$_{d}$ phases, where the
$d$-orbital phases occupy a large part of phase diagrams, indicating large
opportunities for experimental observation. As expected, both SF$_{d}$ and MI$_{d}$ phases appear for a stronger coupling between atoms and cavity mode, as shown in Fig.~\ref{new_detuning_vp}(b). One the other hand, only the SF$_{d}$ phase demonstrates for a weaker atom-cavity coupling, as shown in Fig.~\ref{new_detuning_vp}(a).
%For a weak coupling between atoms and the cavity mode, system supports the superfluid SF$_d$ phase, whereas for a strong coupling, both the $d$-orbital superfluid and Mott-insulating phases appear with increasing the pump lattice depth.

\begin{figure}[tbp]
\includegraphics[width=1\linewidth]{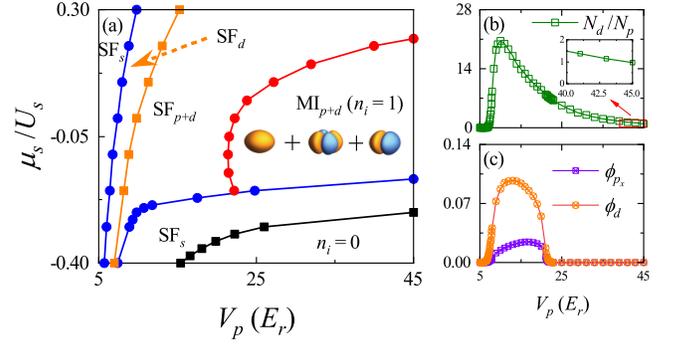}
\caption{Phase diagram of ultracold bosonic gases trapped in an optical cavity, pumped by a blue-detuned laser with an imbalance parameter $\protect\eta =0.8$, obtained from bosonic dynamical mean-field theory. (a) In addition to the $d$-orbital populated phase (SF$_{d}$), we observe superposition of the $p_x$- and $d_{xy}$-orbital atoms in the
self-organized superfluid (SF$_{p+d}$) and Mott-insulating (MI$_{p+d}$) phases. (b) The ratio of total atom number of the $p$- and $d$-orbital bands, and (c) superfluid order parameters $\protect\phi _{p_{x}}$ and $\protect\phi _{d}$, are shown as a function of the pump laser depth, for a fixed chemical potential $\protect\mu %
_{s}/U_{s}=0$. Inset: zoom of the main figure for a stronger
pump strength (b). Other parameters are $\Delta _{c}=100\,\protect\omega _{r}$,
$\protect\kappa =40\,\protect\omega _{r}$, and $N_{\mathrm{lat}}\times
U_{0}=1200\,E_{r}$.}
\label{non_1200}
\end{figure}

\textit{$P$-orbital population for a non-perfect reflection}. In this part, we discuss a non-perfect reflection of pump laser, $\eta <1$. Here, $\hat{V}_{2}\neq 0$ scatters atoms to the $p$-orbital band. This process competes with the $\hat{V}_{1}$
term, which excites atoms to the $d$-orbital band. Therefore, both $p$- and $d$-orbital degrees of freedom come
into play, indicating even richer physics, as shown in Fig.~\ref{non_1200}.%, where we choose the imbalance $\eta =0.8$, cavity detuning $\Delta _{c}=100\, \omega _{r}$, and atom-cavity coupling $N_{\mathrm{% lat}}\times U_{0}=1200\, E_{r}$.

We observe four stable phases, including the SF$_{s}$, SF$_{d}$, $p+d$-orbital superfluid (SF$_{p+d}$), and $p+d$-orbital
Mott-insulating (MI$_{p+d}$) phases. As expected, the system is an $s$-orbital
superfluid phase in the absence of cavity photon for a smaller pumping strength. With{\ the increase of} the pumping power $V_{p}$, more photons
are scattered into the cavity, and atoms organize themselves by being firstly excited to the $d_{xy}$-orbital band, stabilizing the SF$_{d}$ phase. Upon further increasing pumping power, the $p$-orbital band is also populated, and the system enters into a new superfluid phase, SF$_{p+d}$, with both $p_x$- and $d_{xy}$-orbital states being populated. Finally,{\ in the strongly pumping limit}%
, the system enters the Mott-insulating phase, MI$_{p+d}$, with atoms localized in
a superposition of local $p_x$- and $d_{xy}$-orbitals. In this phase, the
self-organized orbital-density wave involves both the $d_{xy}$-orbital $\langle \hat{%
S}_{x}^{sd}\rangle _{i}=-\langle \hat{S}_{x}^{sd}\rangle _{i+1}$, and $p_{x}$%
-orbital $\langle \hat{S}_{x}^{sp_{x}}\rangle _{i}=-\langle \hat{S}%
_{x}^{sp_{x}}\rangle _{i+1}$~\cite{SM}. Note here that only a tiny fraction of atoms populate in the $p_{y}$-orbital state, since scattering atoms to the $p_{y}$-orbital is a higher-order process, as shown in Fig.~\ref{schematic}(d).

To characterize the transition between these phases, population ratio in different bands $N_{d}/N_{p}$, and superfluid order parameters are utilized, as shown in Fig.~\ref{non_1200}(b)(c), where $N_{\sigma
}=\sum_{i}\langle \hat{b}_{i,\sigma }^{\dagger }\hat{b}_{i,\sigma }\rangle $%
. We observe that the population in the $d$-orbital band increases quickly with the pumping strength. However, the population of $p$-orbital is tiny for a shallow lattice, indicating the $d$-orbital phase appearing firstly. When the pumping strength exceeds a critical value, the $p$-orbital band start to be populated, eventually being the same order as $d$-orbital, as shown in the inset of Fig.~\ref{non_1200}(b). Finally, atoms are localized with the absence of superfluid order parameters $\phi_\sigma=0$, indicating the appearance of a Mott-insulating phase, {as shown in Fig.~\ref{non_1200}(c).}

We remark here that the population of higher-orbital states can be tuned by the imbalance $\eta $. When $\eta \sim 1$, ${\hat V}_1$ dominates scattering processes by transferring atoms into the $d$-orbital band with neglected $p$-orbital excitations (Fig.~\ref{new_mu_vp} and~\ref{new_detuning_vp}). For a relatively small $\eta$, the interplay of ${\hat V}_1$ and ${\hat V}_2$ induces superposition of the $p$- and $d$-orbital states (Fig.~\ref{non_1200}). For even smaller $\eta $, ${\hat V}_2$ dominates scattering processes by exciting atoms into the $p$-orbital band, due to the relatively small band gap between the $s$- and $p$-orbital bands. For example, $20\%$ of atoms populate in the $p$-orbital band but with neglected $d$-orbital excitations for $\eta =0.6$, as shown in Fig. S3. Our numerical results thus confirm the possibility for selectively preparing atoms in different higher-orbital bands in an optical cavity system.

\textit{Experimental detections}. For a perfect reflection of pump laser, only $s$- and $d$%
-orbital atoms exist in the self-organized phase, the band-mapping
techniques~\cite%
{PhysRevLett.74.1542,PhysRevLett.87.160405,PhysRevLett.94.080403,PhysRevLett.99.200405}
can be used to distinct atoms in different bands, since the atoms mainly populate
in the quasi-momentum state $\mathbf{k}=(0,0)$ for the $s$-orbital state,
and in $\mathbf{k}=(\pi,\pi)$ for the $d$-orbital one, respectively. But for
the non-perfect reflection, both the $p$- and $d$-orbital atoms mainly populate at the same point $\mathbf{k}=(\pi,\pi)$. In this case, the population in higher-orbital bands can be measured from the images by
non-adiabatically switching off the lattice~\cite%
{PhysRevA.87.063638,PhysRevA.92.043614,Zhou_2018}.

\textit{Conclusions}. An extended Bose-Hubbard model for studying higher-orbital many-body phases is proposed in a cavity setup adapted to recent experiments. By controlling the reflection of the blue-detuned pump laser, we find that atoms can be selectively transferred to the $p$- or $d$-orbital band of a two-dimensional square lattice, and self-organize into stable higher-orbital superfluid and Mott-insulating phases, providing a new mechanics for controlling higher-orbital many-body phenomena. Our current setup only involves square lattice and single-mode cavity. Further work can be extended to complex lattice structures and multiple cavity modes, where orbital frustrations come into play, inducing even richer many-body orbital phenomena.

\textit{Acknowledgements.} This work is supported by the National Natural
Science Foundation of China under Grants No. 12074431, 11304386 and 11774428
(Y.L.), and NSAF No. U1930403 (J.Y.). J.H. is supported by the Postgraduate
Scientific Research Innovation Project of Hunan Province under Grant No.
CX20200012. We acknowledge the Beijing Super Cloud Computing Center (BSCC)
for providing HPC resources that have contributed to the research results
reported within this paper.

\bibliography{references}

\begin{thebibliography}{76}
\bibitem [{lat()}]{lattice_size}%
  \BibitemOpen
  \href@noop {} {}\bibinfo {note} {We consider the system with lattice sites up
  to $N_{\rm lat} = 1024$ and periodic boundary conditions to verify
  finite-size effects on phase diagrams.}\BibitemShut {Stop}%
\end{thebibliography}

\begin{widetext}
	\begin{center}
		{\Large \bf Supplementary Material: Controlling higher-orbital quantum phases of ultracold atoms via
coupling to optical cavities}
	\end{center}
	\renewcommand{\theequation}{S\arabic{equation}}
	\renewcommand{\thesection}{S-\arabic{section}}
	\renewcommand{\thefigure}{S\arabic{figure}}
	\renewcommand{\bibnumfmt}[1]{[S#1]}
	\renewcommand{\citenumfont}[1]{S#1}
	\setcounter{equation}{0}
	\setcounter{figure}{0}

\section{Extended Bose-Hubbard model}
We consider two-level atoms (the ground and excited states are denoted by $|g\rangle$ and $|e\rangle$, respectively) with mass $m$ and transition frequency $\omega_a$, interacting with a single-mode cavity with frequency $\omega_c$ in the $x$ direction, and a pump laser with frequency $\omega_p$ in the $y$ direction. The motion of atoms in the $z$ direction is frozen with strong standing-wave lasers $V_z=50\,E_r$.

We assume two unbalanced counter-propagating beams pumped in the $y$ direction, with the incident light $E_{+}\cos(k_p y-\omega_p t)$ and the reflected light $E_{-}\cos(k_p y+\omega_p t)$ ($E_{-} = \eta E_{+}=\eta E_0$). The total electric field in the $y$ direction can be written as
\begin{eqnarray}
	E(y)&=&E_0\cos(k_p y-\omega_p t) + \eta E_0{\rm cos}(k_p y+\omega_p t)\nonumber\\
	&=&(1+\eta)E_0\cos(k_p y){\rm cos}(\omega_p t) + (1-\eta)E_0\sin(k_p y)\sin(\omega_p t).
\end{eqnarray}
Including the cavity mode in the $x$ direction, the effective interaction between atoms and total electric field in the cavity system is described by
%under rotating-wave and electric-dipole approximation ~\cite{PhysRevLett.95.260401,PhysRevLett.100.050401}
\begin{eqnarray}
	\hat{H}_{\rm int}&=&\hbar\Omega_p(1+\eta){\rm cos}(k_p y)(\hat{\sigma}^+ + \hat{\sigma}^-){\rm cos}(\omega_p t) + \hbar\Omega_p(1-\eta){\rm sin}(k_p y)(\hat{\sigma}^+ + \hat{\sigma}^-){\rm sin}(\omega_p t)\nonumber\\
	&+&\hbar g_0{\rm cos}(k_c x)(\hat{\sigma}^+\hat{a} + \hat{\sigma}^-\hat{a}^\dagger),
\end{eqnarray}
where $\hat{\sigma}^-=|g\rangle\langle e|$, $\hat{\sigma}^+=|e\rangle\langle g|$, $k_{p}$ and $k_{c}$ are wave vectors of the pumping and cavity field, respectively, $\Omega_p$ denotes the maximum pump Rabi frequency, $g_0$ is the atom-cavity coupling strength, and $\hat{a}$ ($\hat{a}^\dagger$) denotes the annihilation (creation) operator of a cavity photon with frequency $\omega_c$. In the reference frame that rotates at the frequency $\omega_p$, this term can be written as
\begin{eqnarray}
	\hat{H}_{\rm int}=\frac{1}{2}\hbar\Omega_p(1+\eta){\rm cos}(k_p y)(\hat{\sigma}^+ + \hat{\sigma}^-)+ \frac{i}{2}\hbar\Omega_p(1-\eta){\rm sin}(k_p y)(\hat{\sigma}^+ - \hat{\sigma}^-)+\hbar g_0{\rm cos}(k_c x)(\hat{\sigma}^+\hat{a} + \hat{\sigma}^-\hat{a}^\dagger).
\end{eqnarray}
Taking the cavity mode and atom degrees of freedom into account, the many-body system can be described by
\begin{eqnarray}\label{single_ham}
	\hat{H}&=&\int d{\bf x} \left[ \hat{\Psi}^\dagger_g ({\bf x})\left(-\frac{\hbar^2\nabla^2}{2m} \right) \hat{\Psi}_g({\bf x}) + \hat{\Psi}^\dagger_e ({\bf x})\left(-\frac{\hbar^2\nabla^2}{2m} - \hbar\Delta_a\right) \hat{\Psi}_e({\bf x})
	\right]- \hbar\Delta_c\hat{a}^\dagger \hat{a}\nonumber\\
	&+&\int d{\bf x} \left[ \hat{\Psi}^\dagger_e ({\bf x})\left(\frac{\hbar\Omega_p(1+\eta){\rm cos}(k_p y)}{2} + \frac{i\hbar\Omega_p(1-\eta){\rm sin}(k_p y)}{2} + \hbar g_0{\rm cos}(k_c x) \hat{a} \right) \hat{\Psi}_g({\bf x}) + {\rm H.c.}
	\right] ,
\end{eqnarray}
where $\hat{\Psi}_g({\bf x})$ ($\hat{\Psi}_e({\bf x})$) denotes the atomic field operator
for annihilating an atom at position ${\bf x}$ in the ground state (excited state).

%And the Heisenberg equation for the excited state is
%\begin{eqnarray}
%	\frac{\partial \hat{\Psi}_e({\bf x})}{\partial t}=i\left[ \left(-\frac{\hbar\nabla^2}{2m} - \Delta_a\right) \hat{\Psi}_e({\bf x})+ \left(\frac{\Omega_p(1+\eta){\rm cos}(ky)}{2} + \frac{i\Omega_p(1-\eta){\rm sin}(ky)}{2}+g_0{\rm cos}(kx) \hat{a}\right)\hat{\Psi}_g({\bf x})
%	\right],
%\end{eqnarray}
%for large detuning $\Delta_a$, we neglect the kinetic energy term and assume that the field operators $\hat{\Psi}_g({\bf x})$ and $\hat{a}$ vary on a much slower time scale than the $1/\Delta_a$ terms, then we arrive at
%\begin{eqnarray}\label{excited}
%	\hat{\Psi}_e({\bf x})=\left(\frac{\Omega_p(1+\eta){\rm cos}(ky)}{2\Delta_a} + \frac{i\Omega_p(1-\eta){\rm sin}(ky)}{2\Delta_a}+\frac{g_0{\rm cos}(kx) \hat{a}}{\Delta_a}\right)\hat{\Psi}_g({\bf x}).
%\end{eqnarray}

Integrating out the excited state of the atom~\cite{PhysRevLett.95.260401,PhysRevLett.100.050401}, the atomic system can be described by an effective Hamiltonian
\begin{eqnarray}\label{eff_system}
\hat{H} &=&  \int d{\bf x}\hat{\Psi}^\dagger({\bf x})\bigg(-\frac{\hbar^2}{2m}\nabla^2 + \eta V_p{\rm cos}^2(k_p y) + U_0{\rm cos}^2(k_c x)\hat{a}^\dagger \hat{a}+\hat{V}_{\mathrm{1,scat}}+\hat{V}_{\mathrm{2,scat}}\bigg)\hat{\Psi}({\bf x})\nonumber\\
&+& \frac g2 \int d{\bf x} \hat{\Psi}^\dagger({\bf x})\hat{\Psi}^\dagger({\bf x})\hat{\Psi}({\bf x})\hat{\Psi}({\bf x})-\hbar\Delta_c \hat{a}^\dagger \hat{a},
\end{eqnarray}
where $\eta=\frac{E_{-}}{E_{+}}$, $\Delta_a = \omega_p - \omega_a$, $\Delta_c = \omega_p - \omega_c$.  $\hat{\Psi}({\bf x})$ denotes the atomic field operator for the ground state, where the excited state has been adiabatically eliminated due to the negligible spontaneous emission for very low temperature $T$ and large detuning $\Delta_c$.
$V_p =\hbar\Omega^2_p/\Delta_a$ is the depth of standing-wave potential created by the pump laser in the $y$ direction, and $U_0=\hbar g^2_0/\Delta_a$ is the light shift of a single maximally
coupled atom. The dominant term is the interference between the pumping and cavity field with $\hat{V}_{\mathrm{1,scat}}=\frac{(1+\eta)}{2} \sqrt{V_p U_0}(\hat{a}+\hat{a}^\dagger) {\rm cos}(k_c x){\rm cos}(k_p y)$, and $\hat{V}_{\mathrm{2,scat}}=-i\frac{(1-\eta)}{2} \sqrt{V_p U_0}(\hat{a}-\hat{a}^\dagger) {\rm cos}(k_c x){\rm sin}(k_p y) $. We remark here that contact interactions $g=\frac{4\pi\hbar^2a_s}{m}$ between atoms have been added in the many-body Hamiltonian, with $a_s$ being the s-wave scattering length. Here, we choose the wave vectors of the pumping and cavity field to be identical with $k_{p}=k_{c}$.
%Here the first term is the kinetic energy of the atoms in the cavity, the second term is the potential of the standing wave formed by the pump laser in the z direction,the third term describes the cavity field with shifted cavity resonance due to the back action of the atoms on the cavity mode, the fourth term $V_1$  and fifth term $V_2$ describes the coherent scattering between the pump laser and the cavity field. And the last term is the interactions between atoms.

Following the standard procedures, the Hamiltonian Eq.~(\ref{eff_system}) can be rewritten in the Wannier basis to obtain an extended Bose-Hubbard model in sufficiently deep lattices. We expand the atomic field operator in the Wannier basis set $\hat{\Psi}({\bf x}) = \sum_{i,\sigma} \hat{b}_{i,\sigma}w_{\sigma}({\bf x}-{\bf x}_i)$, where $\hat{b}_{i,\sigma}$ ($\hat{b}^\dagger_{i,\sigma}$) is the
annihilate (create) operator for a Wannier state $\sigma$ at site $i$, and $w_{\sigma}({\bf x}-{\bf x}_i)$ is the Wannier function centered at ${\bf x} = {\bf x}_i$ for the $s$-, $p_x$-, $p_y$-, and $d_{xy}$-orbital states, respectively. The Bose-Hubbard Hamiltonian with onsite interactions has the following form
%Here, from experimental setup, we define $w_0({\bf x}-{\bf x}_i) = w_0(x)w_0(y)w_0(z)$ and $w_1({\bf x}-{\bf x}_i) = w_1(x)w_1(y)w_1(z)$, thus under tight-binding approximation, the Bose-Hubbard Hamiltonian is given by:
\begin{eqnarray}
\hat{H} =-\sum\limits_{\langle ij\rangle,\sigma }J_{\sigma \sigma }^{ij}\hat{b}_{i,\sigma
}^{\dag }\hat{b}_{j,\sigma }-\sum\limits_{i,\sigma }\mu_{\sigma}\hat{b}_{i,\sigma
}^{\dag }\hat{b}_{j,\sigma }-\hbar \Delta _{c}\hat{a}^{\dagger }\hat{a} + \sum\limits_{i,\sigma _{1}\sigma _{2}\sigma _{3}\sigma _{4}}\frac{%
U_{\sigma _{1}\sigma _{2}\sigma _{3}\sigma _{4}}}{2}\hat{b}_{i,\sigma
_{1}}^{\dag }\hat{b}_{i,\sigma _{2}}^{\dag }\hat{b}_{i,\sigma _{3}}\hat{b}%
_{i,\sigma _{4}} +\hat{%
V}_{1}+\hat{V}_{2},
\end{eqnarray}
where $\hat{V}_{1}=\frac{1+\eta}{2}(\hat{a}+\hat{a}%
^{\dagger })\sum\nolimits_{ij}\left( -1\right) ^{i}(J_{sd}^{ij}\hat{b}%
_{i,s}^{\dag }\hat{b}_{j,d}+J_{p_{x}p_{y}}^{ij}\hat{b}_{i,p_{x}}^{\dag }\hat{%
b}_{j,p_{y}}+\mathrm{H.c.})$, and $\hat{V}_{2}=-i\frac{1-\eta}{2}(\hat{a}-\hat{a}^{\dagger
})\sum\nolimits_{ij}\left( -1\right) ^{i}(J_{sp_{x}}^{ij}\hat{b}_{i,s}^{\dag
}\hat{b}_{j,p_{x}}+J_{p_{y}d}^{ij}\hat{b}_{i,p_{y}}^{\dag }\hat{b}%
_{j,d}+\mathrm{H.c.})$ are the cavity induced scattering processes. Here, $\langle i,j\rangle$ represents the nearest-neighbor sites $i,j$, and the single-particle hopping amplitudes are
\begin{equation}
J^{ij}_{\sigma\sigma} =-\int d \mathbf{x} w^\ast_{\sigma}\left(\mathbf{x}-\mathbf{x}_{i}\right)\left(-\frac{\hbar^2 \nabla^{2}}{2 m} + V_{\rm lat}\right) w_{\sigma}\left(\mathbf{x}-\mathbf{x}_{j}\right)
\end{equation}
\begin{equation}
J^{ij}_{sd}=\int d \mathbf{x}w^\ast_{d_{xy}}\left(\mathbf{x}-\mathbf{x}_{i}\right) \sqrt{V_P U_0} \cos (k_c x) \cos (k_p y) w_{s}\left(\mathbf{x}-\mathbf{x}_{j}\right)
\end{equation}
\begin{equation}
J^{ij}_{p_xp_y}=\int d \mathbf{x}w^\ast_{p_y}\left(\mathbf{x}-\mathbf{x}_{i}\right) \sqrt{V_P U_0} \cos (k_c x) \cos (k_p y) w_{p_x}\left(\mathbf{x}-\mathbf{x}_{j}\right)
\end{equation}
\begin{equation}
J^{ij}_{sp_x}=\int d \mathbf{x}w^\ast_{p_x}\left(\mathbf{x}-\mathbf{x}_{i}\right) \sqrt{V_P U_0} \cos (k_c x) \sin (k_p y) w_{s}\left(\mathbf{x}-\mathbf{x}_{j}\right)
\end{equation}
\begin{equation}
J^{ij}_{p_yd}=\int d \mathbf{x}w^\ast_{d_{xy}}\left(\mathbf{x}-\mathbf{x}_{i}\right) \sqrt{V_P U_0} \cos (k_c x) \sin (k_p y) w_{p_y}\left(\mathbf{x}-\mathbf{x}_{j}\right),
\end{equation}
where $\mu_\sigma \equiv J^{ii}_{\sigma\sigma}$, $V_{\rm lat}\equiv\eta V_p\,{\rm cos}^2(k_py)$ in the pump direction, and $V_{\rm lat}\equiv(V_{\rm cl}+U_0\hat{a}^\dagger\hat{a})\,{\rm cos}^2(k_cx)$ in the cavity direction, with $V_{\rm cl} $ being an external optical lattice added in the cavity direction to validate the tight-binding model. The onsite interaction terms read
\begin{eqnarray}\label{onsite_interaction}
\sum\limits_{i,\sigma _{1}\sigma _{2}\sigma _{3}\sigma _{4}}\frac{%
U_{\sigma _{1}\sigma _{2}\sigma _{3}\sigma _{4}}}{2}\hat{b}_{i,\sigma
_{1}}^{\dag }\hat{b}_{i,\sigma _{2}}^{\dag }\hat{b}_{i,\sigma _{3}}\hat{b}%
_{i,\sigma _{4}}&=&\sum_{i}\bigg(\sum_{\sigma_1\neq\sigma_2\neq\sigma_3\neq\sigma_4}U_{i,\sigma_1\sigma_2\sigma_3\sigma_4}(\hat{b}^\dagger_{i,\sigma_1}\hat{b}^\dagger_{i,\sigma_2}\hat{b}_{i,\sigma_3}\hat{b}_{i,\sigma_4}+{\rm H.c.})\nonumber\\&+&\sum_{\sigma}\frac{U_{i,\sigma}}{2}\hat{n}_{i,\sigma}(\hat{n}_{i,\sigma}-1)+ \sum_{\sigma_1\neq\sigma_2}2U_{i,\sigma_1\sigma_2}\hat{n}_{i,\sigma_1}\hat{n}_{i,\sigma_2}\nonumber\\ &+&\sum_{\sigma_1\neq\sigma_2}\frac{U_{i,\sigma_1\sigma_2}}{2}(\hat{b}^\dagger_{i,\sigma_1}\hat{b}^\dagger_{i,\sigma_1}\hat{b}_{i,\sigma_2}\hat{b}_{i,\sigma_2}+{\rm H.c.})
	\bigg),
\end{eqnarray}
with
\begin{equation}
U_{\sigma _{1}\sigma _{2}\sigma _{3}\sigma _{4}}=\int d \mathbf{x} w^\ast_{\sigma_1}\left(\mathbf{x}-\mathbf{x}_{i}\right) w^\ast_{\sigma_2}\left(\mathbf{x}-\mathbf{x}_{i}\right) \frac{4\pi \hbar^2 a_s}{m} w_{\sigma_3}\left(\mathbf{x}-\mathbf{x}_{i}\right) w_{\sigma_4}\left(\mathbf{x}-\mathbf{x}_{i}\right).
\end{equation}

\begin{figure}
	\includegraphics[width=0.5\linewidth]{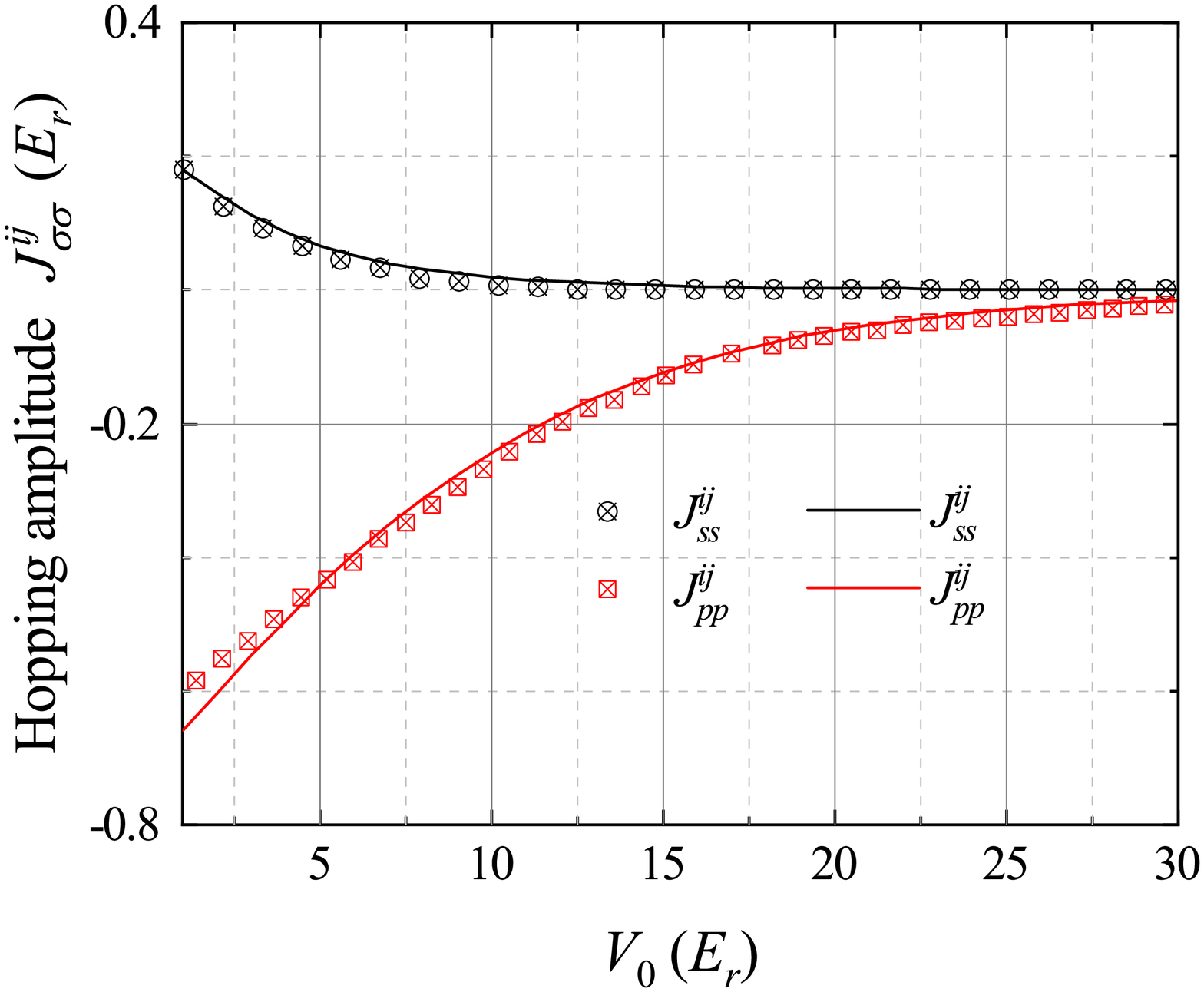}
	\caption{\label{parameter} Nearest-neighbor hopping amplitudes of $J^{ij}_{ss}$ and $J^{ij}_{pp}$ as a function of the lattice depth $V_0\equiv V_x=V_y=V_z$, where the points, denoted by the cross, are from Ref.~\cite{L_hmann_2012}.}
\end{figure}
In order to simplify the effective Hamiltonian, we take the coherent state approximation and represent the cavity mode by a complex amplitude $\alpha$. In the case, the system only depends on the average photon number, and is given by:
\begin{eqnarray}\label{Hamil_2}
&&\hat{H} = - \sum_{\langle ij\rangle,\sigma}J^{ij}_{
	\sigma\sigma}\left(\hat{b}^\dagger_{i,\sigma} \hat{b}_{j,\sigma}+ {\rm H.c.}\right) +\sum\limits_{i,\sigma _{1}\sigma _{2}\sigma _{3}\sigma _{4}}\frac{%
U_{\sigma _{1}\sigma _{2}\sigma _{3}\sigma _{4}}}{2}\hat{b}_{i,\sigma
_{1}}^{\dag }\hat{b}_{i,\sigma _{2}}^{\dag }\hat{b}_{i,\sigma _{3}}\hat{b}%
_{i,\sigma _{4}} -\sum_{i,\sigma}\mu_{\sigma}\hat{b}^\dagger_{i,\sigma} \hat{b}_{i,\sigma}  - \hbar |\alpha|^2 \Delta_c + Re[\alpha](1+\eta) \nonumber \\
&&\,\,\,\,
 \sum_{ij}(-1)^{i}\left(J^{ij}_{sd} \hat{b}^\dagger_{i,s} \hat{b}_{j,d} + J^{ij}_{p_xp_y} \hat{b}^\dagger_{i,p_x} \hat{b}_{j,p_y} + \mathrm{H.c.} \right) +Im[\alpha](1-\eta)\sum_{ij}(-1)^{i} \left( J^{ij}_{sp_x} \hat{b}^\dagger_{i,s} \hat{b}_{j,p_x} + J^{ij}_{p_yd}\hat{b}^\dagger_{i,p_y}\hat{b}_{j,d} + \mathrm{H.c.}  \right),
\end{eqnarray}
%where $J_{sd}$, $J_{sp_x}$, $J_{p_xp_y}$ and $J_{p_yd}$ denote the onsite scattering between the cavity mode and the pump laser, $J^\prime_{sd}$, $J^\prime_{sp_x}$, $J^\prime_{p_xp_y}$ and $J^\prime_{p_yd}$ are the hopping amplitudes induced by the scattering between the cavity mode and pump field.
where $\alpha =\frac{\sum_{i}(-1)^{i}\langle (1+\eta) (J^{ii}_{sd}%
\hat{b}_{i,s}^{\dagger }\hat{b}_{i,d}+J^{ii}_{p_{x}p_{y}}\hat{b}%
_{i,p_{x}}^{\dagger }\hat{b}_{i,p_{y}}+\mathrm{H.c.})+i(1-\eta)(J^{ii}_{sp_{x}}\hat{b}%
_{i,s}^{\dagger }\hat{b}_{i,p_{x}}+J^{ii}_{p_{y}d}\hat{b}_{i,p_{y}}^{\dagger }%
\hat{b}_{i,d}+\mathrm{H.c.})\rangle}{2(\Delta _{c}-\sum_{i,\sigma
}J_{\sigma }\langle \hat{b}_{i,\sigma }^{\dagger }\hat{b}_{i,\sigma }\rangle
+i\kappa )}$, with $J_{\sigma} =\int d \mathbf{x} w^\ast_{\sigma}\left(\mathbf{x}-\mathbf{x}_{i}\right)U_{\rm 0}\,{\rm cos}^2(k_cx) w_{\sigma}\left(\mathbf{x}-\mathbf{x}_{i}\right)$, and $\kappa$ being the decay rate of the cavity mode. We remark here that all the Hubbard parameters are obtained from the band-structure simulations, where the nearest-neighbor hopping amplitudes for the lowest two bands are compared with the data from Ref.~\cite{L_hmann_2012}, as shown in Fig.~\ref{parameter}. Actually, in a deep optical lattice, the localized wavefunctions can be described as the product of two
Wannier functions for each direction, $w_{s}({\bf x-x}_i)= w_{s}(x)w_{s}(y)$, $w_{p_x}({\bf x-x}_i)=w_{p}(x)w_{s}(y)$, $w_{p_y}({\bf x-x}_i)=w_{s}(x)w_{p}(y)$, and $w_{d_{xy}}({\bf x-x}_i)=w_{p}(x)w_{p}(y)$, where $w_s(x)$ and $w_p(x)$ \big($w_s(y)$ and $w_p(y)$\big) denote the Wannier functions of the lowest two Bloch bands of a one-dimensional lattice in the $x$ ($y$) direction, respectively.

%Note that, all the zero-order tunneling terms irrelevant to the cavity field are absorbed by the global chemical potential. %where $J_{\sigma}$ denotes the onsite single-particle matrix elements of the potential generated by the cavity mode for site $i$. $\alpha = \frac {\sum_{i} \langle J_{sd} \hat{b}^\dagger_{s}\hat{b}_{i,d} +J_{p_xp_y} \hat{b}^\dagger_{p_x}\hat{b}_{i,p_y}+ {\rm H.c.} + i \left(J_{sp_x}\hat{b}^\dagger_{i,s}\hat{b}_{i,p_x}+J_{p_yd}\hat{b}^\dagger_{i,p_y}\hat{b}_{i,d}+ {\rm H.c.}\right)\rangle} {\Delta_c - \sum_{i,\sigma}\langle J_{\sigma}\hat{b}^\dagger_{i,\sigma} \hat{b}_{i,\sigma} \rangle + i\kappa}$,

\section{bosonic dynamical mean-field theory}
\subsection{BDMFT equations}
We derive a self-consistent equations within bosonic dynamical mean-field theory (BDMFT) by using the cavity method~\cite{Appen_georges96}, which is suitable for the high but finite dimensional optical lattice. The effective action of the impurity site up to subleading order in $1/z$ is described in the standard way~\cite{Appen_georges96, Vollhardt}
\begin{eqnarray}\label{eff_action}
	S^{(0)}_{imp}&=&\int_{0}^{\beta} d\tau d\tau'\sum_{\sigma_1,\sigma_1',\sigma_2, \sigma_2'}\left(
	\begin{array}{c} b^*_{0,\sigma_1}(\tau)\\
		b_{0,\sigma_1}(\tau)
	\end{array}
	\right)^{T}\mathcal{G}^{-1}_{0,\sigma_1\sigma_2\sigma_1'\sigma_2'}(\tau-\tau')
	\left(
	\begin{array}{c} b_{0,\sigma_2}(\tau')\\
		b^*_{0,\sigma_2}(\tau')
	\end{array}
	\right)\nonumber \\
&-&\int_{0}^{\beta} d\tau \sum_{\langle 0j \rangle, \sigma_1,\sigma_1^\prime}(-1)^{\delta_{\sigma_1\sigma_1^\prime}+1}J^{0j}_{\sigma_1\sigma_1^\prime}[b^\ast_{0,\sigma_1}(\tau) \phi_{j,\sigma_1'}(\tau)+{\rm H.c.}]
	\nonumber\\
	&+&\int_{0}^{\beta} d\tau\left( \sum_{\sigma_1,\sigma_1^\prime}J^{00}_{\sigma_1\sigma_1^\prime}b^\ast_{0,\sigma_1}(\tau)b_{0,\sigma_1^\prime}(\tau)+{\rm H.c.}+\frac12\sum_{\sigma_1\sigma_2\sigma_3\sigma_4}U_{\sigma_1\sigma_2\sigma_3\sigma_4}b^{(0)\ast}_{\sigma_1}(\tau)b^{(0)\ast}_{\sigma_2}(\tau)b^{(0)}_{ \sigma_3}(\tau)b^{(0)}_{\sigma_4}(\tau)\right),
\end{eqnarray}
where $J^{00}_{\sigma_1\sigma_1'}$ denotes the onsite hopping amplitudes induced by pump-cavity scattering, and $J^{0j}_{\sigma_1\sigma_1'}$ is nearest-neighbor hopping amplitudes induced by the kinetic energy and pump-cavity scattering. Note here that the site-dependent parameters $(-1)^iRe[\alpha](1+\eta)$ and $(-1)^i Im[\alpha](1-\eta)$ are absorbed by the scattering induced hopping amplitudes, to shorten the effective action. The Weiss Green's function ($8\times8$ matrix) is defined as
\begin{eqnarray}
	&&\mathcal{G}^{-1}_{0,\sigma_1\sigma_2\sigma_1'\sigma_2'}(\tau-\tau')=\\
	&&\left(
	\begin{array}{cc}
		 (\partial_{\tau'}-\mu_{\sigma_1})\delta_{\sigma_1\sigma_2}+\sum_{\langle 0j \rangle,\langle 0j' \rangle}J^{0j}_{\sigma_1\sigma_1'}J^{0j'}_{\sigma_2\sigma_2'}G^1_{j,j',\sigma_1',\sigma_2'}(\tau,\tau') & \sum_{\langle 0j \rangle,\langle 0j' \rangle}J^{0j}_{\sigma_1\sigma_1'}J^{0j'}_{\sigma_2\sigma_2'}G^2_{j,j',\sigma_1',\sigma_2'}(\tau,\tau')\\
		\sum_{\langle 0j \rangle,\langle 0j' \rangle}J^{0j}_{\sigma_1\sigma_1'}J^{0j'}_{\sigma_2\sigma_2'}G^{*2}_{j,j',\sigma_1',\sigma_2'}(\tau',\tau) &(-\partial_{\tau'}-\mu_{\sigma_1})\delta_{\sigma_1\sigma_2}+\sum_{\langle 0j \rangle,\langle 0j' \rangle}J^{0j}_{\sigma_1\sigma_1'}J^{0j'}_{\sigma_2\sigma_2'}G^{1}_{j,j',\sigma_1',\sigma_2'}(\tau',\tau)
	\end{array}
	\right)\nonumber
\end{eqnarray}
and we introduce
\begin{equation}
\phi^{}_{j,\sigma_1}(\tau) \equiv \langle b_{j, \sigma_1} (\tau)
\rangle_0
\end{equation}
as the superfluid order parameters, and
\begin{eqnarray}
&&G^1_{j,j',\sigma_1',\sigma_2'}(\tau,\tau')=\langle b_{j,\sigma_1'}(\tau) b^*_{j',\sigma_2^\prime}(\tau')\rangle_{(0)}-  \phi_{j,\sigma_1'}(\tau) \phi^*_{j',\sigma_2^\prime}(\tau')\\
&&G^2_{j,j',\sigma_1',\sigma_2'}(\tau,\tau')=\langle b_{j,\sigma_1'}(\tau) b_{j',\sigma_2'}(\tau')\rangle_{(0)}-  \phi_{j,\sigma_1'}(\tau) \phi_{j',\sigma_2'}(\tau')
\end{eqnarray}
as the diagonal and off-diagonal parts of the connected Green's functions, respectively. $\langle \ldots \rangle_0$ takes the expectation value in the cavity system excluding the impurity site.

\subsection{Anderson impurity model}
It's difficult to find a  solver analytically for the effective action Eq.~(\ref{eff_action}). Therefore, we turn back to the Hamiltonian representation to obtain BDMFT equations. The effective action Eq.~(\ref{eff_action}), can be represented by the Anderson impurity Hamiltonian
\begin{eqnarray}
\hat{H}^{(0)}_A &=& - \sum_{\sigma} J^{0j}_{\sigma\sigma} \Bigg( \Big(\phi^{(0)*}_{\sigma} \hat{b}^{(0)}_{\sigma} + {\rm H.c.} \Big) - \mu_{\sigma} \hat{n}^{(0)}_{\sigma} \Bigg)+ \frac12\sum_{\sigma_1\sigma_2\sigma_3\sigma_4}U_{\sigma_1\sigma_2\sigma_3\sigma_4}b^{(0)\ast}_{\sigma_1}(\tau)b^{(0)\ast}_{\sigma_2}(\tau)b^{(0)}_{ \sigma_3}(\tau)b^{(0)}_{\sigma_4}(\tau) \nonumber  \\
&+& \left(J^{0j}_{sd}\hat{b}_{s}^{(0)*} \phi_{d}^{(0)} + J^{0j}_{p_xp_y} \hat{b}_{p_x}^{(0)*} \phi_{p_y}^{(0)}+{\rm H.c.} \right) +  \left(J^{0j}_{sp_x}\hat{b}_{s}^{(0)*} \phi_{p_x}^{(0)} + J^{0j}_{p_yd}\hat{b}_{p_y}^{(0)*} \phi_{d}^{(0)} +{\rm H.c.} \right)\nonumber\\
&+&\left(J^{00}_{sd} \hat{b}_{s}^{(0)*} \hat{b}_{d}^{(0)}  +J^{00}_{p_xp_y} \hat{b}_{p_x}^{(0)*} \hat{b}_{p_y}^{(0)}+{\rm H.c.}\right)+ \left(J^{00}_{sp_x}\hat{b}_{s}^{(0)*} \hat{b}_{p_x}^{(0)}+J^{00}_{p_yd}\hat{b}_{p_y}^{(0)*} \hat{b}_{d}^{(0)} +{\rm H.c.} \right)  \nonumber \\
&+& \sum_{l}  \epsilon_l \hat{a}^\dagger_l\hat{a}_l + \sum_{l,\sigma} \Big( V_{\sigma,l} \hat{a}^\dagger_l\hat{b}^{(0)}_{\sigma} + W_{\sigma,l} \hat{a}_l\hat{b}^{(0)}_{\sigma} + {\rm H.c.} \Big),
\end{eqnarray}
where the the onsite terms including chemical potential, interaction and onsite scattering terms are directly inherited from the Hubbard Hamiltonian. BDMFT couples two different baths, where the condensed bath of bosons is represented by the Gutzwiller term with superfluid order parameters $\phi^{(0)}_{\sigma}$ for the $s$-, $p_x$-, $p_y$- and $d_{xy}$-orbital states. The normal bath of bosons is described by a finite number of orbitals with creation operators $\hat{a}^\dagger_l$ and energies
$\epsilon_l$, where these orbitals are coupled to the impurity via normal-hopping amplitudes
$V_{\sigma, l}$ and anomalous-hopping amplitudes $W_{\sigma, l}$, which are needed to generate the off-diagonal elements of the hybridization functions.

To obtain the solution of the impurity model, the Anderson Hamiltonian is straightforwardly implemented in the Fock basis, and the corresponding solution can be achieved by exact diagonalization of dynamical mean-field theory ~\cite{Appen_georges96,Hubener}. %Details can be found in~\cite{Vollhardt, Hubener, Werner,Li2011,Li2012,Li2013, Liang15, Li2016, PhysRevLett.121.093401}.

\begin{figure}
	\includegraphics[width=0.7\linewidth]{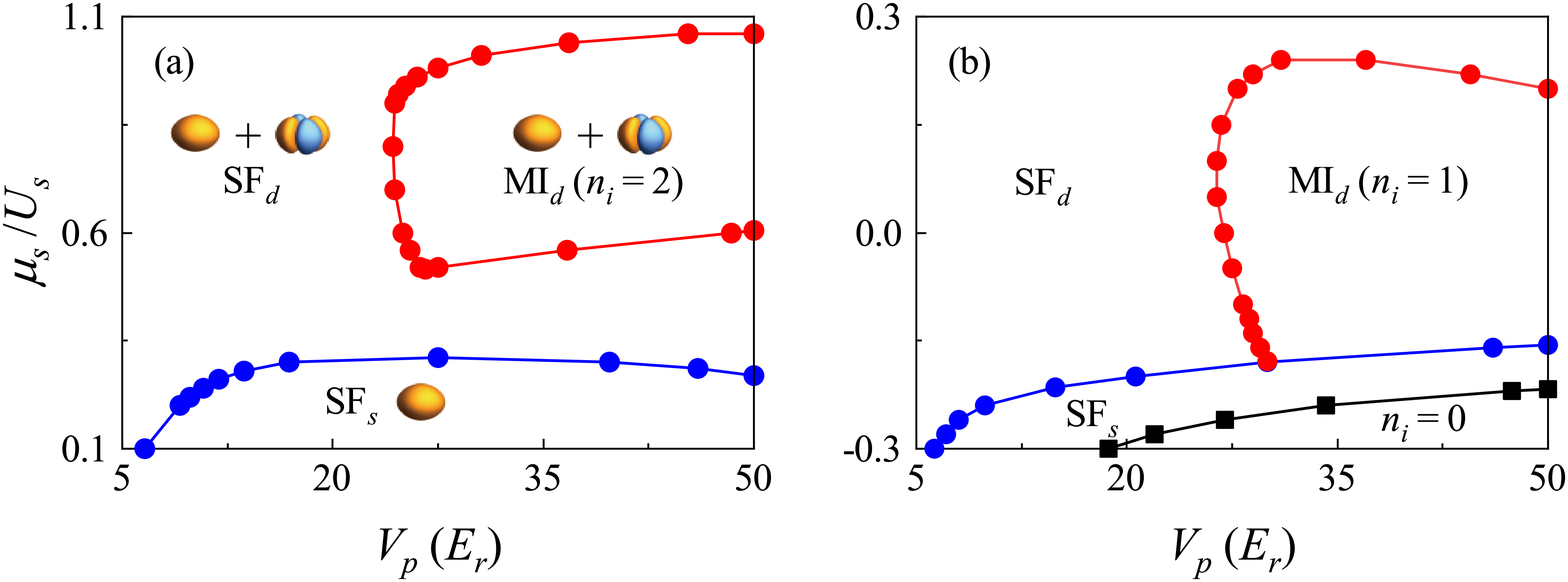}
	\caption{\label{mu_vp} Phase diagram of ultracold bosonic gases trapped in a cavity, pumped by a blue-detuned laser with the imbalance parameter $\eta=1$ for different atom-cavity couplings $N_{\rm{lat}}\times U_0 = 500\,E_r$ (a), and $1200\,E_r$ in (b), obtained from bosonic dynamical mean-field theory. Here, there are three many-body phases, including the $s$-orbital superfluid phase (SF$_s$) in the absence of photon in the cavity, $d$-orbital superfluid phase (SF$_d$) in the presence of photons in the cavity, and $d$-orbital Mott-insulating phase (MI$_d$) with localized $d$-orbital atoms, respectively. Other parameters are $\Delta_c=80\,\omega_r$, and $\kappa=40\,\omega_r$.}
\end{figure}
\section{Phase diagrams for different Hubbard parameters}
\subsection{Phase diagram for $\eta=1$}
For the system with a perfect reflection with $\eta=1$, we observe various interesting phases in the blue-detuned system, which are stable for a large parameter regime. Actually, there is an intermediate regime for the coupling $N_{\rm lat}\times U_0$ where the scattering processes become pronounced, in contrast to the red-detuned case with larger coupling favoring self-organization of atoms. In this section, we discuss these phases in different regimes for the coupling $N_{\rm lat}\times U_0$. As shown in Fig.~\ref{mu_vp}(a) and (b), there are also three many-body phases, including the $s$-orbital superfluid phase (SF$_s$) in the absence of photon in the cavity, $d$-orbital superfluid phase (SF$_d$) in the presence of photons in the cavity, and $d$-orbital Mott-insulating phase (MI$_d$) with localized $d$-orbital atoms.  % Here, $N_{\rm{lat}}\times U_0=1200E_r$ in (a) and $N_{\rm{lat}}\times U_0=500E_r$ in (b), respectively. For a larger imbalance, $\eta\approx 1$, atoms are mainly scattered from the $s$- to $d$-orbital band of a square lattice, but with tiny $p$-orbital population, as shown in Fig. 2 and 3 in the main text.

\subsection{Phase diagram for $\eta<1$}
The ratio of the population in the $p$- and $d$-orbital bands can tuned by the reflection of the pump laser. For a relatively small imbalance, such as $\eta=0.8$, atoms are scattered from the $s$-orbital to $p$- and $d$-orbital bands of a two-dimensional square lattice, as shown in Fig. 4 in the main text. Upon decreasing the imbalance parameter $\eta$, it is expected that more and more atoms are scattered into the $p$-orbital band, due to the relatively small band gap between the $s$- and $p$-orbital bands. Eventually, atoms scattering into the $d$-orbital band can be neglected, and $p$-orbital scattering processes dominate instead. For example, we find that atoms are dominantly scattered into the $p$-orbital band but with neglected excitations in the $d$-orbital band for $\eta =0.6$, as shown in Fig.~\ref{new_non_1200}(b), where a new phase, the $p$-orbital superfluid phase (SF$_p$), is found, defined as $| \alpha|^2\neq0$ and $\phi_p \neq0$.  As shown in Fig.~\ref{new_non_1200}(a), there are two many-body phases, including the $s$-orbital superfluid phase (SF$_s$) in the absence of photon in the cavity, and $p$-orbital superfluid phase (SF$_p$) in the presence of photons in the cavity.

\begin{figure}
	\includegraphics[width=0.5\linewidth]{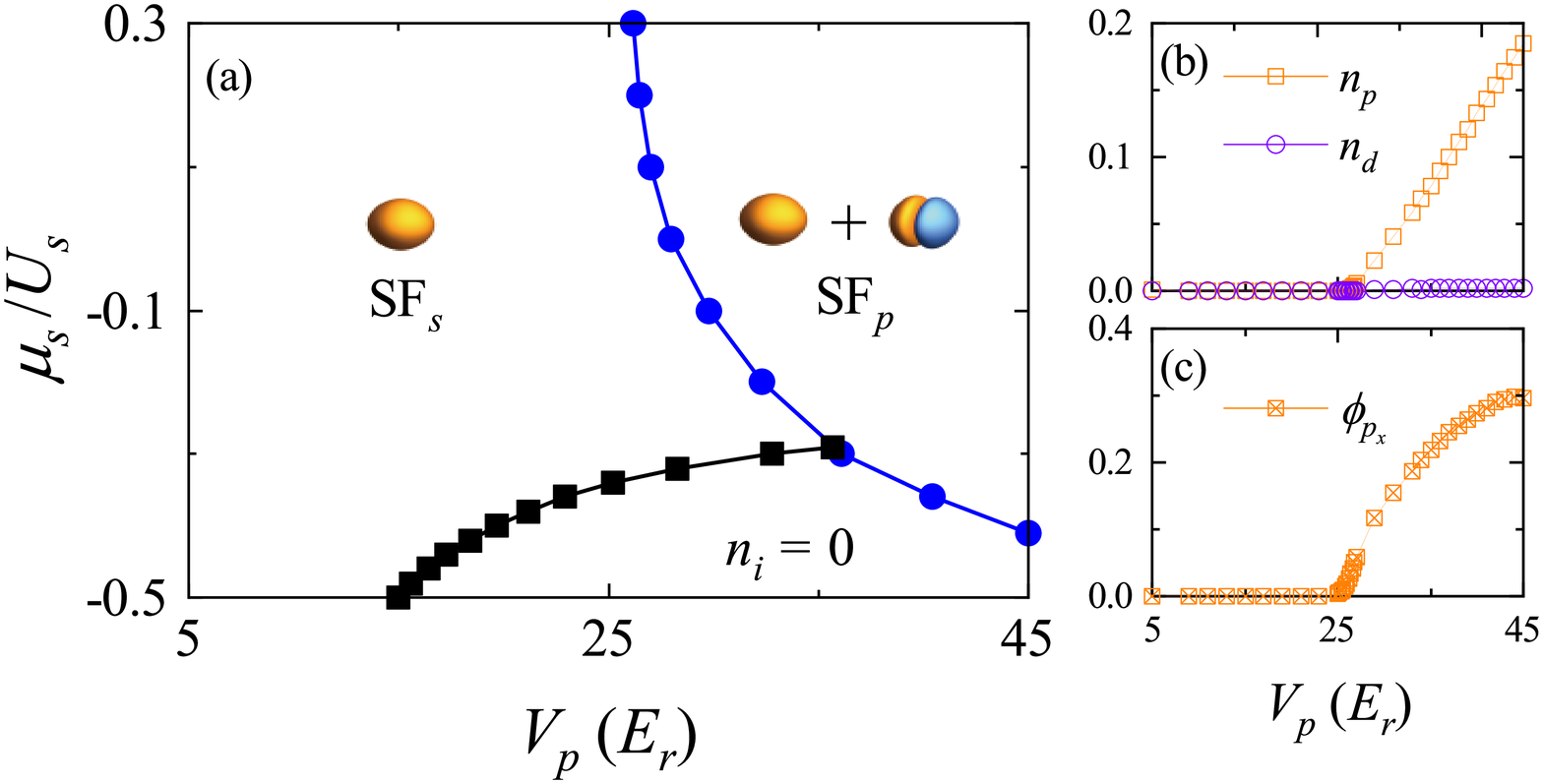}
	\caption{\label{new_non_1200} (a) Phase diagram of ultracold bosonic gases trapped in an optical cavity, pumped by a blue-detuned laser with a non-perfect reflection $\eta=0.6$. (b) Averaged population of the atoms in the $p$-orbital states $n_p=\sum_i (n_{i,p_x} + n_{i,p_y})/N_{\rm lat}$ and the $d$-orbital state $ n_d=\sum_i n_{i,d}/N_{\rm lat}$, and (c) order parameter $\phi_{p_x}$ are shown as a function of the pump laser depth along the line $\mu_s=0.3U_s$, indicating scattering processes involved the $p$-orbital states being dominant. The other parameters are $N_{\rm{lat}}\times U_0 = 1200\,E_r$, and $\Delta_c=-20\,\omega_r$.}
\end{figure}

\section{Self-organized orbital-density wave order in the deep Mott-insulating regime}
Orbital degrees of freedom can be treated as pseudo-spins, and one obtains self-organized orbital-density wave orders. For example, one treats the $s$-orbital atoms as spin $\uparrow$ and the $d$-orbital ($p$-orbital) atoms as spin $\downarrow$, and essentially achieves a pseudospin-1/2 system in optical lattices. Here, local orbital magnetism of the system is given by
${\bf \hat{S}}^{\sigma_1\sigma_2} = \hat{b}^\dagger_{i,\sigma} {\bf F}_{\sigma\sigma'} \hat{b}_{i,\sigma'} $,
with ${\bf F}_{\sigma\sigma^\prime}$ denoting the spin matrix for a spin-1/2 particle, {\it i.e.} $\hat{S}^{sd}_x= 1/2  ({\hat{b}_{i,s}}^\dagger {\hat{b}_{i,d}} + {\hat{b}_{i,d}}^\dagger {\hat{b}_{i,s}}) $, $\hat{S}^{sd}_y=i/2 (-{ \hat{b}_{i,s}}^\dagger { \hat{b}_{i,d}} + {\hat{b}_{i,d}}^\dagger {\hat{b}_{i,s}}) $, and $\hat{S}^{sd}_z= 1/2  ({\hat{b}_{i,s}}^\dagger {\hat{b}_{i,s}} - {\hat{b}_{i,d}}^\dagger  {\hat{b}_{i,d}} )$ for the $s$- and $d$-orbital degrees of freedom.
Similarly, $\hat{S}^{sp_x}_x= 1/2  ({\hat{b}_{i,s}}^\dagger {\hat{b}_{i,p_x}} + {\hat{b}_{i,p_x}}^\dagger {\hat{b}_{i,s}}) $, $\hat{S}^{sp_x}_y=i/2 (-{\hat{b}_{i,s}}^\dagger {\hat{b}_{i,p_x}} + {\hat{b}_{i,p_x}}^\dagger {\hat{b}_{i,s}}) $, and $\hat{S}^{sp_x}_z= 1/2  ({\hat{b}_{i,s}}^\dagger {\hat{b}_{i,s}} - {\hat{b}_{i,p_x}}^\dagger  {\hat{b}_{i,p_x}} )$ for the $s$- and $p_x$-orbital degrees of freedom.

\subsection{Orbital-density wave order for filling $n_i=1$ and reflection $\eta=1$}
For a perfect reflection of the blue-detuned pump laser with $\eta=1$, the atoms are transferred from the $s$- to $d$-orbital state with neglected populations in the $p$-orbital band. Therefore, we eliminate the terms related to $p$-orbital degrees of freedom, and rewrite the Eq.~(\ref{Hamil_2}) in the zero-hopping limit
\begin{eqnarray}
	\hat{H}_0 &=& Re[\alpha](1+\eta)\sum_i (-1)^{i}J^{ii}_{sd} \left(\hat{b}^\dagger_{i,s} \hat{b}_{i,d} + {\rm H.c.}  \right) +\frac{U_s}{2}\sum_i\hat{n}_{i,s}(\hat{n}_{i,s}-1) + \frac{U_d}{2}\sum_i\hat{n}_{i,d}(\hat{n}_{i,d}-1) + 2U_{sd}\hat{n}_{i,s}\hat{n}_{i,d}\nonumber\\
	&+& \frac{U_{sd}}{2}\left(\hat{b}^\dagger_{i,s}\hat{b}^\dagger_{i,s}\hat{b}_{i,d}\hat{b}_{i,d}+{\rm H.c.} \right)+ \sum_{i,\sigma}\mu_{\sigma}\hat{b}^\dagger_{i,\sigma} \hat{b}_{i,\sigma},
\end{eqnarray}
where $\hat{H}_0$ includes the onsite interactions, scattering terms and chemical potential.

In the deep Mott-insulating phase with local total filling $n_i=1$, the single-site Hamiltonian for site $i$ can be written in a matrix form under the basis of $\vert s \rangle$ and $\vert d \rangle$
\begin{eqnarray}
	H_i=\left(\begin{array}{cc} \mu_s & J_1\cdot(-1)^i\\
		J_1\cdot(-1)^i & \mu_d\\
	\end{array}\right),
\end{eqnarray}
where $J_1=Re[\alpha](1+\eta)J^{ii}_{sd}$, and the states $|s\rangle={\hat b}^\dagger_{i,s}|0\rangle$ and $|d\rangle={\hat b}^\dagger_{i,d_{xy}}|0\rangle$ for site $i$.

After diagonalizing the Hamiltonian $H_i$, we obtain the eigenstates and eignenergies,
\begin{eqnarray}
	\vert g_1\rangle&&=(-1)^{i+1}\frac{\mu_d-\mu_s+A_1}{\sqrt{(\mu_d-\mu_s+A_1)^2+4{J_1}^2}}\vert s \rangle+\frac{2J_1}{\sqrt{(\mu_d-\mu_s+A_1)^2+4{J_1}^2}}\vert d \rangle, E_{g_1}=\frac{1}{2}(\mu_d+\mu_s-A_1)\\
	\vert g_2\rangle&&=(-1)^{i+1}\frac{\mu_d-\mu_s-A_1}{\sqrt{(\mu_d-\mu_s-A_1)^2+4{J_1}^2}}\vert s \rangle+\frac{2J_1}{\sqrt{(\mu_d-\mu_s-A_1)^2+4{J_1}^2}}\vert d \rangle, E_{g_2}=\frac{1}{2}(\mu_d+\mu_s+A_1),
\end{eqnarray}
where $A_1=\sqrt{4J_1^2+\mu_d^2-2\mu_d\mu_s+\mu_s^2}$. The $J_1$ term is normally a big positive value in the deep $d$-orbital Mott phase, indicating that $\vert g_1\rangle$ being the ground state and the corresponding orbital-density wave order being,
\begin{eqnarray}\label{S_mag}
	\langle \hat{S}^{sd}_x \rangle_i = -\langle \hat{S}^{sd}_x \rangle_{i+1},
\end{eqnarray}
where $\langle \cdots \rangle_i$ denotes the average value for site $i$.
%We also calculate the long-range order using the real parameters obtained from band-structure simulations, and find that it does not change the final conclusion in the deep $d$-orbital Mott-insulating regime. where the parameters are obtained from the ground and excited states of the two-dimensional harmonic oscillator

\subsection{Orbital-density wave order for filling $n_i=2$ and reflection $\eta=1$}
We now extend the discussion to the case of two atoms per site $n_i=2$ in the deep Mott-insulating phase. We consider the case with interactions $U\equiv U_{sd}\approx\frac{1}{4}U_s\approx \frac{1}{2}U_d$, where the ratio of the parameters are obtained from band-structure simulations in the deep Mott-insulating regime. In the basis of $|s,s\rangle$, $|s,d\rangle$ and $|d,d\rangle$, the single-site Hamiltonian for site $i$ can be written as
\begin{eqnarray}
	H_i=\left(\begin{array}{ccc}
		2\mu_s+4U & J_1\cdot(-1)^i & U\\
		J_1\cdot(-1)^i & \mu_s+\mu_d+2U & J_1\cdot(-1)^i\\
		U & J_1\cdot(-1)^i & 2\mu_d+2U
	\end{array}\right),
\end{eqnarray}
with $J_1=Re[\alpha](1+\eta)J^{ii}_{sd}$. After diagonalizing the Hamiltonian, the eigenstates and eigenvalues are given by
\begin{eqnarray}
	\vert g_1\rangle&=&-\frac{J_1}{A_2}\vert s, s \rangle+(-1)^{i+1}\frac{\mu_d-\mu_s-U}{A_2}\vert s, d \rangle+\frac{J_1}{A_2}\vert d,d\rangle, E_{g_1}=\mu_d+\mu_s+2U\\
	\vert g_2\rangle&=&-\frac{\mu_s-\ \mu_d+2U-D_2}{B_2}\vert s, s \rangle+(-1)^{i+1}\frac{2J_1}{B_2}\vert s, d \rangle+\frac{\mu_s-\mu_d+D_2}{B_2}\vert d, d \rangle, E_{g_2}=\mu_d+\mu_s+3U-D_2\\
	\vert g_3\rangle&=&-\frac{\mu_d-\ \mu_s-2U-D_2}{C_2}\vert s, s \rangle+(-1)^{i}\frac{2J_1}{C_2}\vert s, d \rangle+\frac{\mu_d-\mu_s+D_2}{C_2}\vert d, d \rangle, E_{g_3}=\mu_d+\mu_s+3U+D_2,
\end{eqnarray}
with $A_2=\sqrt{2J_i^2+(\mu_d-\mu_s-U)^2}$, $B_2=\sqrt{(\mu_s-\mu_d+2U-D_2)^2+4J_1^2+(\mu_s-\mu_d+D_2)^2}$, $C_2=\sqrt{(\mu_d-\mu_s-2U-D_2)^2+4J_1^2+(\mu_d-\mu_s+D_2)^2}$, and $D_2=\sqrt{2J_1^2+(\mu_d-U)^2+(\mu_s+U)^2-2\mu_d\mu_s}$. One also obtains $\langle \hat{S}^{sd}_x \rangle_i = -\langle \hat{S}^{sd}_x \rangle_{i+1}$ under the ground state $\vert g_2\rangle$ for $U-D_2<0$ in deep $d$-orbital Mott-insulating phase.
%We remark here that onsite interactions and chemical potential obtained from band simulations are normally different, but this do not influence the final conclusion of the orbital order of the cavity system.

\subsection{Orbital-density wave order for filling $n_i=1$ and reflection $\eta<1$}
For a non-perfect reflection of the pump laser, the field in the pumping direction is not an idea standing wave, and has some running-wave component. In this case, the  $s$-orbital atoms can be excited to both the $p$- and $d$-orbital states. In the absence of nearest-neighbor hopping terms, the Hamiltonian is given by
\begin{eqnarray}
	\hat{H}_0 &=&  Re[\alpha](1+\eta)\sum_i (-1)^{i}\left(J^{ii}_{sd} \hat{b}^\dagger_{i,s} \hat{b}_{i,d}  +J^{ii}_{p_xp_y} \hat{b}^\dagger_{i,p_x} \hat{b}_{i,p_y} + {\rm H.c.}\right)\nonumber\\
&+&Im[\alpha](1-\eta)\sum_i (-1)^{i} \left(J^{ii}_{sp_x}\hat{b}^\dagger_{i,s} \hat{b}_{i,p_x}+J^{ii}_{p_yd}\hat{b}^\dagger_{i,p_y} \hat{b}_{i,d} +{\rm H.c.} \right)  \nonumber \\
	 &+&\frac12\sum_{i,\sigma,\sigma^\prime,\sigma^{\prime\prime},\sigma^{\prime\prime\prime}}U_{\sigma\sigma^\prime\sigma^{\prime\prime}\sigma^{\prime\prime\prime}}\hat{b}^\dagger_{i,\sigma}\hat{b}^\dagger_{i,\sigma^\prime}\hat{b}_{i, \sigma^{\prime\prime}}\hat{b}_{i,\sigma^{\prime\prime\prime}} +\sum_{i,\sigma}\mu_{\sigma}\hat{b}^\dagger_{i,\sigma} \hat{b}_{i,\sigma}.
\end{eqnarray}

We focus on the deep Mott-insulating phase with filling $n_i=1$. In the basis of $|s\rangle$, $|p_x\rangle$ and $|d\rangle$, the single-site Hamiltonian for site $i$ can be written as
\begin{eqnarray}
	H_i=\left(\begin{array}{ccc}
		\mu & J_2\cdot(-1)^i & J_1\cdot(-1)^i\\
		J_2\cdot(-1)^i & \mu & 0 \\
		J_1\cdot(-1)^i & 0 & \mu
	\end{array}\right),
\end{eqnarray}
where $\mu=\mu_{\sigma}$, $J_1=Re[\alpha](1+\eta)J^{ii}_{sd}$, and $J_2=Im[\alpha](1-\eta)J^{ii}_{sp_x}$. Note here that the $p_y$-orbital state has been neglected, since the $p$-orbital population is tiny in our numerical simulations.

After diagonalize the Hamiltonian, the eigenstates with eigenenergies are given by
\begin{eqnarray}
	\vert g_1\rangle&=&-\frac{1}{\sqrt{B_3^2+1}}\vert p \rangle+\frac{B_3}{\sqrt{B_3^2+1}}\vert d \rangle, E_{g_1}=\mu\\
	\vert g_2\rangle&=&(-1)^{i+1}\frac{A_3}{\sqrt{A_3^2+B_3^2+1}}\vert s\rangle+\frac{B_3}{\sqrt{A_3^2+B_3^2+1}}\vert p \rangle+\frac{1}{\sqrt{A_3^2+B_3^2+1}}\vert d \rangle, E_{g_2}=\mu-\sqrt{J^2_1+J^2_2}\\
	\vert g_3\rangle&=&(-1)^{i}\frac{A_3}{\sqrt{A_3^2+B_3^2+1}}\vert s\rangle+\frac{B_3}{\sqrt{A_3^2+B_3^2+1}}\vert p \rangle+\frac{1}{\sqrt{A_3^2+B_3^2+1}}\vert d \rangle, E_{g_3}=\mu+\sqrt{J^2_1+J^2_2},
\end{eqnarray}
with $A_3=\sqrt{J^2_1+J^2_2}/J_1$, and $B_3=J_2/J_1$. The long-range orders of the ground state $\vert g_2\rangle$ are given by
\begin{eqnarray}\label{S_mag_1}
	\langle \hat{S}^{sp_x}_x \rangle_i = -\langle \hat{S}^{sp_x}_x \rangle_{i+1},
\end{eqnarray}
and
\begin{eqnarray}\label{S_mag_2}
	\langle \hat{S}^{sd}_x \rangle_i = -\langle \hat{S}^{sd}_x \rangle_{i+1}.
\end{eqnarray}
We also check the long-range orders using the real parameters obtained via band-structure simulations, and find similar orbital-density wave patterns, as shown above.

%By calculating the ground states for the system with perfect or non-perfect reflection of the pump laser, the self-organized phase with spin waves due to the onsite staggered scattering process between $s$-orbital and $p$-orbital($d$-orbital) can be understood well.

\clearpage

%\begin{thebibliography}{99}
%\bibitem{Ho_98} T. L. Ho, Phys. Rev. Lett. {\bf 81}, 742 (1998).
%\bibitem{Ham_two_spin}  M. Luo, Z.-B. Li, and C.-G. Bao, Phys. Rev. A {\bf 75}, 043609 (2007); Z.-F. Xu Y.-B. Zhang and L. You, Phys. Rev. A {\bf 79}, 023613 (2009).
%\bibitem{DJWang15_appen}  X.-K. Li, B. Zhu, X.-D. He, F.-D. Wang, M.-Y. Guo, Z.-F. Xu, S.-Z. Zhang, and D.-J. Wang, Phys. Rev. Lett. \textbf{114}, 255301 (2015).
%\bibitem{single_mode_98} C. K. Law, H. Pu, and N. P. Bigelow, Phys. Rev. Lett. {\bf 81}, 5257 (1998).
%\bibitem{georges96} A. Georges, G. Kotliar, W. Krauth and M. J. Rozenberg, Rev. Mod. Phys. {\bf 68}, 13 (1996).
%\bibitem{Byczuk_2008} K. Byczuk and D. Vollhardt, Phys. Rev. B {\bf 77}, 235106 (2008).
%\bibitem{M. Caffarel_1994} M. Caffarel and W. Krauth, Phys. Rev. Lett. {\bf 72}, 1545 (1994).
%\bibitem{Walter} M. Snoek and W. Hofstetter, {\it Bosonic Dynamical Mean-Field Theory} Chapter in {\it "Quantum Gases: Finite Temperature and Non-Equilibrium Dynamics"}, N.P. Proukakis {\it et al.}, (Imperial College Press, London, 2013).
%\end{thebibliography}

\end{widetext}

\end{document}